\documentclass[twocolumn,aps,showpacs,prb,tightenlines,amsmath,amssymb]{revtex4}
\usepackage{bm}
\usepackage{graphicx}              
\usepackage{amssymb}               
\usepackage{amsmath}                     
\usepackage{colordvi}
\usepackage{calrsfs} 
\newcommand{\bgreek}[1]{\mbox{\boldmath$#1$\unboldmath}}
\begin{document}   

\title{Anomalous D'yakonov-Perel' spin relaxation in InAs (110) 
quantum wells under strong magnetic field: role of Hartree-Fock self-energy}
\author{T. Yu}
\author{M. W. Wu}
\thanks{Author to whom correspondence should be addressed}
\email{mwwu@ustc.edu.cn}
\affiliation{Hefei National Laboratory for Physical Sciences at
  Microscale and Department of Physics, 
University of Science and Technology of China, Hefei,
  Anhui, 230026, China} 
\date{\today}

\begin{abstract} 
We investigate the influence of the Hartree-Fock self-energy, acting as an
effective magnetic field, on the anomalous
D'yakonov-Perel' spin relaxation in InAs (110) quantum wells when the magnetic
field in the Voigt configuration is much stronger than the
spin-orbit-coupled field. The transverse and longitudinal spin relaxations
 are discussed both analytically and numerically. For the
transverse configuration, it is found that the spin relaxation is very sensitive
to the Hartree-Fock effective magnetic
field, which is very different from the conventional D'yakonov-Perel' spin relaxation. Even an extremely
small spin polarization ($P=0.1\%$) can significantly influence the behavior of
the spin relaxation. It is further revealed that this comes from the {\em unique} form
of the effective inhomogeneous broadening, originated from the mutually perpendicular spin-orbit-coupled field and
  strong magnetic field. It is shown that this effective inhomogeneous
  broadening is very small and hence very sensitive to the
  Hartree-Fock field. Moreover, we further find that in the spin polarization
  dependence, the transverse spin relaxation time decreases with
the increase of the spin polarization in
the intermediate spin polarization regime, which is also very different from the
conventional situation, where the spin relaxation is always suppressed
by the Hartree-Fock field. It is revealed that this {\em opposite} trends come from the additional spin
relaxation channel induced by the HF field. For the
longitudinal configuration, we find that the spin relaxation can be either suppressed or enhanced by the
Hartree-Fock field if the spin polarization is parallel or
antiparallel to the magnetic field. 

\end{abstract}
%72.25.Rb 	Spin relaxation and scattering 
%71.70.Ej 	Spin-orbit coupling, Zeeman and Stark splitting, Jahn-Teller
%effect 
%73.21.Fg 	Quantum wells (73.21.-b 	Electron states and
%collective excitations in multilayers, quantum wells, mesoscopic, and
%nanoscale systems (for electron states in nanoscale materials, see
%73.22.-f)
\maketitle

 \section{Introduction}
For the sake of potential spintronic application, the spin dynamics in
semiconductor quantum wells (QWs) has been extensively studied in the past
decades.\cite{review1,Bloch,Awschalom,Zutic,fabian565,Dyakonov,wu-review,Korn,notebook}
Among these studies, the spin relaxation is one of the most
important problems. In $n$-type semiconductor QWs, it has been understood that
the spin relaxation is limited by the D'yakonov-Perel' (DP) mechanism.\cite{DP,Bloch,wu-review} 
Very recently, it has been further reported by Zhou {\it et al.} that in the framework
of the DP mechanism, the spin relaxation time (SRT) $\tau_s$ and the momentum
relaxation time
$\tau_p^{\ast}$ show anomalous scalings in InAs (110) QWs under a strong
magnetic field ${\bf B}$ in the Voigt configuration satisfying $|g|\mu_BB\gg \langle
|\Omega_z({\bf k})|\rangle$ ($g$ is the effective Land\'e
factor).\cite{Zhou} There, the effective magnetic field reads ($\hbar \equiv 1$ throughout this paper)\cite{Ohno110,Wu110,Dohrmann04,spin_noise110,Sherman110,Zhou_previous}
\begin{equation}
{\bf{\Omega}}({\bf{k}})=\big(\Omega,0,\gamma_Dk_x(k_x^2-2k_y^2-\langle
k_z^2\rangle)/2\big),
\label{110}
\end{equation} 
with $\gamma_D$, ${\bf
  k}=(k_x,k_y)$ and $\Omega\equiv g\mu_B B$ standing for the Dresselhaus
coefficient,\cite{dres} electron momentum and 
Zeeman magnetic field, respectively. Due to the unique form of the effective magnetic
field, by varying the impurity density, the transverse (longitudinal) spin
relaxation can be divided into four (two) regimes: the normal weak
scattering regime ($\tau_{\rm s}\propto \tau_p^{\ast}$), the anomalous DP-like
regime ($\tau_{\rm s}^{-1}\propto \tau_p^{\ast}$),
the anomalous Elliott-Yafet- (EY-)like\cite{Elliott,Yafet}  regime ($\tau_{\rm s}\propto \tau_p^{\ast}$)
and the normal strong scattering regime ($\tau_{\rm s}^{-1}\propto
\tau_p^{\ast}$) (the anomalous EY-like regime and the normal strong scattering
regime).\cite{Zhou} 

In the work of Zhou {\it et al.}, the spin
polarization $P$ is {\em extremely}
small ($P=0.1\%$ in their work) and the predicted spin relaxations can
only be measured by the spin-noise spectroscopy.\cite{Sherman4,Sherman5}
However, this extremely small spin polarization can be hardly resolved by optical orientations, such as the
photo-luminescence\cite{Bloch,photoluminescence1,photoluminescence2} and Faraday/Kerr
measurements.\cite{Faraday_Kerr,Awschalom,wu-review,Aws_rotation2,Aws_rotation3,Korn}
 Therefore, it is essential to study the anomalous DP spin
relaxation beyond the extremely small spin polarization. When larger spin
polarization is considered, it was theoretically
predicted\cite{wu-review,jianhua23}
 and then experimentally verified\cite{jianhua36,jianhua37,jianhua46} that
 the Hartree-Fock (HF) self-energy, acting as
an effective magnetic field, 
\begin{equation}
{\bf \Omega}_{\rm HF}({\bf k})=-\sum_{{\bf k'}}V_{{\bf
    k-k'}}\mbox{Tr}[\rho_{\bf k'}\bgreek \sigma],
\label{HF_field}
\end{equation}
with $V_{{\bf k-k'}}$ being the screened
Coulomb potential, can efficiently suppress the conventional DP spin relaxation.
 Similarly, here in the anomalous DP spin
relaxation, the HF effective magnetic
 field is expected to cause rich and intriguing physics. Specifically, by noting that under the strong magnetic
field, in the
  anomalous DP-like regime, the effective inhomogeneous broadening\cite{2001} is extremely
  small (about $0.1\%$ as large as the one in the absence of the strong
    magnetic field),\cite{Zhou} even a small spin polarization ($P\approx 0.1\%$ as revealed
later) and hence a weak HF effective magnetic field can significantly
influence the transverse spin relaxation. This marked suppression of the
  spin relaxation by the HF magnetic field with such small spin polarization
  $P\approx 0.1\%$ is beyond expectation and has not yet been reported in the
 literature.\cite{jianhua36,jianhua37,jianhua46}

In the present work,
 we utilize the kinetic spin Bloch equations (KSBEs) to study the role of
   the HF effective magnetic field to the anomalous DP spin
 relaxation in InAs (110) QWs.\cite{wu-review}  With the HF effective magnetic field
[Eq.~(\ref{HF_field})] explicitly included, when $\langle|\Omega_{\rm HF}({\bf k})|\rangle\ll
|\Omega|$, the
transverse and longitudinal SRTs due to the electron-impurity scattering
 in the strong scattering limit [$\langle \Omega_{\rm eff}({\bf
  k})\rangle\tau_{k,2}\ll 1$] become, \cite{Yu} 
\begin{equation}
\tau_{{\rm s}z}^{-1}=\left\langle\Omega^2_{\rm eff}({\bf k})\tau_{k,2}+\frac{\overline{{\Omega_z({\bf k})}}^2\tau_{k,1}}{2(1+\Omega^2\tau_{k,1}^2)}\right\rangle
\label{zz_relaxation}
\end{equation}
and
\begin{eqnarray}
\tau_{{\rm s}x}^{-1}&=&\left\langle\frac{{\overline{\Omega_z({\bf
    k})}}^2\tau_{k,1}}{1+[\Omega+\Omega_{\rm HF}({\bf k})]^2\tau_{k,1}^2}\right\rangle,
\label{xx_relaxation}
\end{eqnarray} 
respectively, with the effective inhomogeneous broadening being
\begin{equation} 
\Omega_{\rm eff}({\bf k})\equiv{|\overline{\Omega^2_z({\bf
      k})}|}/(2\Omega\sqrt{1+\Omega^2_{\rm HF}({\bf k})\tau_{k,2}^2}).
\label{effective}
\end{equation}
Here, $\overline{A_{\bf k}} = A_{\bf k}-\frac{1}{2\pi}\int d \phi_{\bf
    k}\,A_{\bf k}$
and the momentum relaxation time $\tau_{k,l}$ limited by the electron-impurity scattering reads
\begin{equation}
\tau_{k,l}^{-1}=\frac{N_i}{2\pi}\int_0^{2\pi}d\phi_{\bf k-k'}|V_{\bf
  k-k'}|^2(1-\cos l\phi_{\bf k-k'}),
\end{equation}
with $N_i$ being the impurity density.

These results are consistent with our recent work
  in the three-dimensional spin-orbit-coupled ultracold Fermi gas, with similar effective magnetic
  fieldcreated by the Raman beams.\cite{Yu,Wang}  However, in the two
  dimensional electron gas (2DEG) system,
new features arise due to the {\em k-dependent} Coulomb potential, which is
different from the contact potential, i.e., constant $V$ in the cold atom
system.\cite{interaction,interaction2,pwave1,Spielman,Yu} With the Coulomb
potential, the HF effective
magnetic field [Eq.~(\ref{HF_field})] is also {\em k-dependent} and not exactly along the direction of the
spin polarization (In the cold atom system, ${\bf \Omega}_{\rm HF}$ is always
antiparallel to the spin polarization\cite{Yu}). Moreover, in the cold atom system,  both the momentum
relaxation time ($\tau_p^{*}\propto V^{-2}$) and the HF effective magnetic field
($\Omega_{\rm HF}\propto V$) vary simultaneously when tuning the interatom
interaction potential by the Feshbach resonance.\cite{Feshbach} Here, in the 2DEG system, the scattering
strength and the HF effective magnetic field can be tuned {\em separately} by varying the impurity density and the spin
polarization. This provides a platform to study the effects
of the scattering and the HF effective magnetic field to the anomalous DP spin
relaxation, separately. Furthermore, by noting that in cold atoms, when tuning the interatom interaction
potential, $\Omega_{\rm HF}^2\tau_{p}^*$ is fixed, in the scattering
potential dependence, the transverse
SRT contributed by the effective inhomogeneous
broadening [Eq.~({\ref{effective}})] in the anomalous DP-like regime when $|\Omega_{\rm
  HF}|\tau_p^*\gg 1$  is independent of the scattering potential.
 Here, we can utilize the 2DEG system to study the contribution of the effective inhomogeneous
broadening in the anomalous DP-like regime to the transverse SRT explicitly.

In this work, we find that for the transverse configuration,
 the effective inhomogeneous broadening is very small and can be significantly
 suppressed by the HF field. Hence the spin relaxation is very
sensitive to the HF field and even a small spin polarization can
significantly influence the transverse spin relaxation. Specifically, with
$P=0.1\%$, the normal strong scattering regime in the absence of the HF field
vanishes and hence the {\em whole} scattering is in the strong scattering limit.
Interestingly, a slightly further increase of the spin polarization with
$P\gtrsim 1\%$, the transverse SRT is
  divided into two rather than four regimes: the anomalous EY-like regime and the normal strong scattering
regime. Moreover, it is found that in the intermediate spin polarization
regime, the HF field can provide an additional spin relaxation channel, and
hence the
transverse SRT decreases with the increase of the spin polarization. On the other hand,
the longitudinal spin relaxation is again divided into the anomalous EY and normal strong scattering
regimes in the presence of the HF field. It is further observed that the
longitudinal spin relaxation can be either suppressed or enhanced by the
HF effective magnetic field if the spin polarization is parallel or
antiparallel to the magnetic field.

This paper is organized as follows. In Sec.~II, we present the main results. In
Sec.~{\ref{weak_part}},
 we study the influence of the weak HF effective magnetic field [$\langle|\Omega_{\rm HF}({\bf
k})|\rangle\ll |\Omega|$] on the spin
 relaxation in the impurity density
dependence of the SRT with different spin polarizations. We compare the analytical and numerical results in detail with
only the electron-impurity scattering. We
also discuss the situation with all the relevant scatterings. In Sec.~{\ref{strong_part}}, we
study the anomalous spin relaxation under the strong HF effective magnetic field
[$\langle|\Omega_{\rm HF}({\bf
k})|\rangle> |\Omega|$] and present the spin polarization dependence of the spin relaxation. We
summarize in Sec.~{\ref{summary}}.

\section{Results}
The KSBEs are written as\cite{wu-review}
\begin{equation}
  \partial_t \rho_{\bf k}(t)=\partial_t\rho_{\bf k}(t)|_{\rm coh}+\partial_t\rho_{\bf k}(t)|_{\rm  scat},
\label{ksbe}
\end{equation}
in which $\rho_{\bf k}(t)$ represent the density matrices of
electrons with momentum ${\bf k}$  at
 time $t$. The coherent terms $\partial_t\rho_{\bf k}(t)|_{\rm
   coh}$ describe the spin precessions of electrons due to the effective magnetic
 field ${\bf \Omega(k)}$ and the HF self-energy. The scattering terms $\partial_t\rho_{\bf k}(t)|_{\rm
  scat}$ include the electron-impurity, electron-electron and electron-phonon
scatterings. The expressions for the coherent and scattering terms
can be found in Ref.~\onlinecite{Zhoujun}.

In the numerical calculation, $g=-14.3$
(Ref.~\onlinecite{g-factor}) and $\gamma_D=-27.3$~eV~\AA$^3$.\cite{Kotani_gamma_D} The other material parameters can
be found in Ref.~\onlinecite{Jianhua}. The temperature and the electron density are set to be $T=30$ K and
$N_e=3\times 10^{11}$ cm$^{-2}$, respectively. We further set $B=4$ T, with the condition $|\Omega|\gg \langle
|\Omega_z({\bf k})|\rangle$ satisfied and the
Zeeman splitting energy being much smaller than the Fermi energy. Moreover, the well width is set to be $a=5$ nm,
which is much smaller than the cyclotron radius of the lowest Landau
level.

With these material parameters, by varying the initial
spin polarization and hence the HF effective magnetic field
[Eq.~(\ref{HF_field})], both situations with weak HF effective magnetic field $\langle|\Omega_{\rm HF}({\bf
k})|\rangle\ll |\Omega|$ and strong one $\langle|\Omega_{\rm HF}({\bf k})|\rangle\gtrsim
|\Omega|$ can be realized. In Secs.~{\ref{weak_part}} and B, we discuss the spin
relaxation with weak HF effective magnetic field $\langle|\Omega_{\rm HF}({\bf
k})|\rangle\ll |\Omega|$ and strong one $\langle|\Omega_{\rm HF}({\bf
k})|\rangle\gtrsim |\Omega|$, respectively.

\subsection{Weak HF effective magnetic field}
\label{weak_part} 
\begin{figure}[ht]
  {\includegraphics[width=7.4cm]{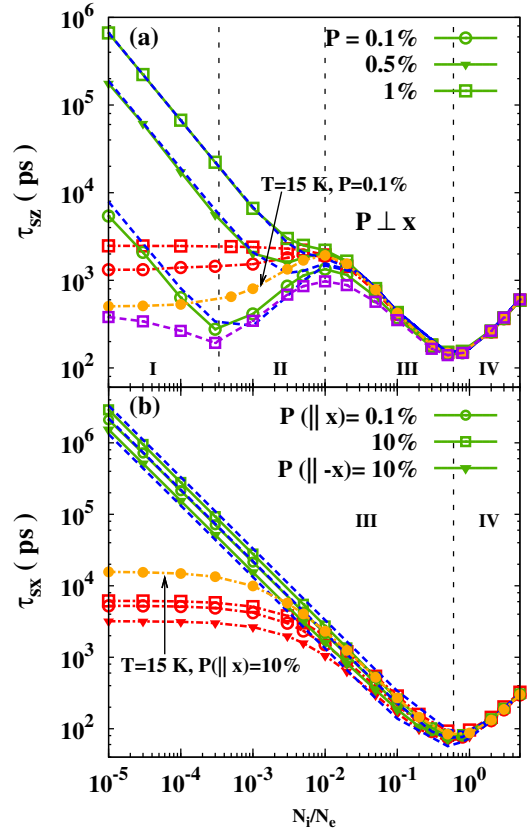}}
  \caption{(Color online) Transverse (a) and longitudinal (b)
SRTs with only the electron-impurity scattering
(green solid curves) and all the relevant scatterings (red chain curves) against the
impurity density with different initial spin polarizations. In (a) [(b)], the
blue dashed curves represent the analytical results calculated from Eq.~(\ref{zz_relaxation})
[Eq.~(\ref{xx_relaxation})]. In the calculation, the temperature is set to be 30 K, except for the case plotted by the orange chain curve with dots is calulated at 15 K, representing the
transverse (longitudinal) SRT with all the relevant scatterings when $P=0.1\%$ ($10\%$). The purple dashed curve with squares in (a) represents the case with $P=0.1\%$ in the absence of the HF field. The Roman numbers represent different regimes of the spin
  relaxation: I, the normal weak scattering regime; II, the anomalous DP-like regime; III,
the anomalous EY-like regime; IV, the normal strong scattering regime. The vertical
black dashed lines indicate the boundaries between different
regimes for the transverse (longitudinal) spin relaxation with $P=0.1\%$ in the absence (presence) of the HF field.}
  \label{figyw1}
\end{figure}

In this subsection, we discuss the spin relaxation with weak HF effective
magnetic field [$\langle|\Omega_{\rm HF}({\bf k})|\rangle\ll
|\Omega|$]. In Fig.~\ref{figyw1}, we plot the transverse
and longitudinal SRTs against the
impurity density with different initial spin polarizations. Below we first
analyze the transverse and longitudinal SRTs with only the electron-impurity scattering and
compare the numerical results with the analytical ones; then we discuss the
genuine situation with all the relevant scatterings.

\subsubsection{Impurity scattering}
In this part, we analyze the spin relaxation with only the electron-impurity
scattering.
 With weak HF effective magnetic field [$\langle|\Omega_{\rm HF}({\bf k})|\rangle\ll
|\Omega|$], for the transverse (longitudinal) spin relaxation, the HF effective
magnetic field
 acts as a rotating (static)
magnetic field around (along) the Zeeman field.\cite{Yu} In this situation,
for the electron-impurity scattering, which is elastic, the transverse
(longitudinal) SRT is expressed in Eq.~(\ref{zz_relaxation})
[Eq.~(\ref{xx_relaxation})]. We first compare the numerical results with the
analytical ones from Eqs.~(\ref{zz_relaxation}) and (\ref{xx_relaxation}). In
Fig.~\ref{figyw1}, the numerical results and the analytical ones are plotted
as the green solid and blue dashed curves with different spin polarizations. It
is shown that the numerical results agree fairly well with the analytical ones
in the {\em whole} scattering regime for the transverse [Fig.~\ref{figyw1} (a)] and
longitudinal [Fig.~\ref{figyw1} (b)] SRTs. Accordingly, the underlying
physics of the transverse (longitudinal) spin relaxation with weak HF effective magnetic
field can be understood facilitated with Eq.~(\ref{zz_relaxation})
[Eq.~(\ref{xx_relaxation})].

 We first analyze the transverse spin relaxation [Fig.~\ref{figyw1}(a)]. For
 reference, we present the SRT with the extremely small spin polarization
($P=0.1\%$) in the absence of the HF effective magnetic field (shown by the 
purple dashed curve with squares). It is shown that the SRTs in this situation can be
divided into four regimes:
 I, the normal weak scattering regime; II, the anomalous DP-like
regime; III, the anomalous EY-like regime; IV, the normal strong scattering
regime (the boundaries are indicated by the vertical
black dashed lines).\cite{Zhou} When the HF effective magnetic field is included
for the situation with $P=0.1\%$ ($0.5\%$),  the transverse spin relaxation {\em seemingly} can still be divided
into four regimes, with the SRT in Regimes I and II enhanced. 
 When the spin polarizations are slightly
 beyond the extremely small spin polarization ($P=0.1\%$), it is shown in
 Fig.~\ref{figyw1}(a) that with $P=1\%$, the SRT is divided into two rather than four
 regimes. One also observes that with the increase of the initial spin polarization, the
anomalous DP-like regime gradually disappears and the SRTs in Regimes I and
II are significantly enhanced; whereas the SRTs in Regimes III and IV remain
unchanged.  These marked influences of such small spin polarizations
  ($P=0.1\%$ and $1\%$) on the spin relaxation are well beyond
  expectation [In conventional situation, the effects of 
the HF effective magnetic field become obvious only when $P\gtrsim
  10\%$ (Refs.~\onlinecite{jianhua36,jianhua37,jianhua46})].
 Below we show that these
intriguing phenomena arise from the suppression of the {\em unique} effective
inhomogeneous broadening
 [Eq.~(\ref{effective})]
by the HF effective magnetic field.

We begin the analysis from the case with $P=0.1\%$.  When the HF effective
 magnetic field is included,
 the SRT is still divided into four regimes, but the underlying physics is
very different from the case without the HF field. One notes that when $P=0.1\%$, in Regime I (II), $\langle |\Omega_{\rm
HF}({\bf k})|\tau_{k,2}\rangle\gtrsim 1$ [$\langle |\Omega_{\rm
HF}({\bf k})|\tau_{k,2}\rangle\ll 1$] is satisfied. Accordingly, in Regime I, when $\langle|\Omega_{\rm HF}({\bf k})|\rangle\gtrsim \Omega_{\rm HF}^c$ [$\langle|\Omega_{\rm HF}({\bf k})|\rangle\ll \Omega_{\rm HF}^c$] with
\begin{equation}
\Omega_{\rm HF}^c\equiv\langle|\overline{\Omega^2_z({\bf k})}
|\rangle/{(2|\Omega|)},
\label{critical}
\end{equation}
$\langle\Omega_{\rm eff}({\bf k})\rangle\tau_{k,2}\approx 
\langle|\overline{\Omega^2_z({\bf
    k})}|/|2\Omega\Omega_{\rm HF}({\bf k})|\rangle$ is smaller (larger)
 than 1, and hence the normal weak scattering regime is absent (present). By noting
 that $\Omega_{\rm HF}^c$ is extremely small due to the strong Zeeman
 magnetic field, even with extremely small spin polarization $P=0.1\%$, the HF
 effective magnetic field can exceed $\Omega_{\rm HF}^c$ and the normal weak
 scattering regime vanishes. That is to say, when $P=0.1\%$, the {\em whole}
 scattering regime is in the strong scattering limit and all the behaviors of the
 transverse SRT 
can be analyzed facilitated with Eq.~(\ref{zz_relaxation}).

  Specifically, in Rigme I with $\langle |\Omega_{\rm
HF}({\bf k})|\tau_{k,2}\rangle\gg 1$, the SRT 
\begin{equation}
\tau_{sz}\approx\left\langle 4\Omega^2\Omega_{\rm HF}({\bf
  k})\tau_{k,2}\Big{/}\overline{\Omega_z^2({\bf k})}^2\right\rangle 
\end{equation} 
is proportional to the momentum relaxation time, showing the EY-like
behavior (Without the HF effective magnetic field, this behavior is understood due to the spin relaxation channel provided by the momentum
relaxation in the weak scattering limit\cite{Zhou}). In Regime II with $\langle |\Omega_{\rm
HF}({\bf k})|\tau_{k,2}\rangle\ll 1$, the SRT is still inversely proportional to
the momentum relaxation time, and can be
enhanced due to the suppression of the effective inhomogeneous broadening
[Eq.~(\ref{effective})] by the HF effective magnetic field. In Regimes III and
IV, the SRT is uninfluenced by the HF effective magnetic field. Furthermore, the
boundary between Regimes I/II, II/III and III/IV can be determined from Eq.~(\ref{zz_relaxation}).   
The position of the basin at the
 crossover between I/II is
 determined by
 \begin{equation}
 \tau_{k,2}\approx \langle|\Omega_{\rm HF}({\bf k})|\rangle^{-1},
 \label{boundary1}
 \end{equation}
(instead of $\tau_{k,2}\approx \langle\Omega_{\rm eff}({\bf k})\rangle^{-1}$ in the
absence of the HF effective magnetic field\cite{Zhou}); the position of the
 peak at the crossover between II/III is determined by
 \begin{equation}
 \tau_{k,1}\tau_{k,2}\approx\left\langle\overline{\Omega_z({\bf
     k})}^2\Big{/}[2\Omega^2\Omega_{\rm eff}^2({\bf k})]\right\rangle;
 \label{boundary2}
 \end{equation}
the basin at the crossover between III/IV is determined by
\begin{equation}
\tau_{k,1}\approx|\Omega|^{-1}.
\end{equation}

One notes that from Eq.~(\ref{boundary1}) [Eq.~(\ref{boundary2})], when the HF
effective magnetic field increases, the boundary between Regimes I/II (II/III) is shifted to the
stronger (weaker) scattering. Hence with larger spin polarization
$P=0.5\%$, the range of Regime II is significantly
suppressed. Furthermore, one expects that with the spin polarization, and hence the HF
effective magnetic field large enough, Regime II may vanish. This phenomenon
arises in our situation with $P=1\%$. When $P=1\%$, in Regimes I and II, by noting that $\langle|\Omega_{\rm HF}({\bf
  k})|\rangle\tau_{k,2}\gtrsim 1$ is satisfied, the SRT becomes 
\begin{equation}
\tau_{{\rm s}z}\approx\left\langle \overline{\Omega^2_z({\bf k})}^2\Big{/}
[4\Omega^2\Omega^2_{\rm HF}({\bf k})\tau_{k,2}]\right\rangle^{-1},
\end{equation}
which shows the EY-like behavior. 
With the original anomalous EY-like and normal strong scattering
 regimes keeping unchanged, the spin relaxation is divided into two regimes: the anomalous 
EY-like and normal strong scattering regimes.

It is noted that when $P\lesssim 0.02\%$, the condition $\langle|\Omega_{\rm HF}({\bf k})|\rangle\ll \Omega_{\rm HF}^c$ is satisfied, and the normal weak scattering regime still arises. The underlying physics then returns to the one revealed in Ref.~\onlinecite{Zhou} with the HF effective magnetic field absent.

Although weak HF magnetic field can have marked influence on the transverse spin
relaxation, for the longitudinal configuration, the situation is very
different. Below we analyze the longitudinal spin relaxation [Fig.~\ref{figyw1}(b)].  For reference, the SRT with
an extremely small spin polarization ($P=0.1\%$) is plotted, which can be
divided into two regimes: III, the anomalous EY-like regime
 and IV, the normal strong scattering regime.\cite{Zhou} 
 When the spin polarization is large with $P=10\%$, the
spin relaxation also presents similar behavior to the situation with small spin
polarization, showing a basin with the increase of the impurity
density. Furthermore, compared with the situation with $P=0.1\%$, when the
initial spin
polarization is  parallel (antiparallel) to the external magnetic field, the SRT
is {\em slightly} enhanced (suppressed) in the anomalous EY-like regime and unchanged in the normal strong
scattering regime. This is shown by the green solid curve with squares
(triangles) in Fig.~\ref{figyw1}(b) for the spin polarization parallel (antiparallel) to the
external magnetic field. These phenomena can be understood as follows.

 With the weak HF effective magnetic field, from Eq.~(\ref{xx_relaxation}), we
 conclude that the longitudinal SRT is divided into two regimes with the
 boundary determined by 
\begin{equation}
 \tau_{k,1}=\langle|\Omega+\Omega_{\rm HF}({\bf k})|\rangle^{-1}.
 \end{equation}
Furthermore, one notes that the HF effective magnetic field is parallel to the spin
 polarization
 [Eqs.~(\ref{HF_field})]. Hence in the anomalous EY-like
 regime, with $\langle|\Omega_{\rm HF}({\bf k})|\rangle\ll |\Omega|$, when the spin
polarization is parallel (antiparallel) to the external magnetic
field, the total magnetic field $|\Omega+\Omega_{\rm HF}({\bf k})|$ along the
$\hat{x}$-axis is {\em slightly} enhanced (suppressed) by the HF effective magnetic
field. Accordingly,
 from Eq.~(\ref{xx_relaxation}), in which $|\Omega+\Omega_{\rm HF}({\bf k})|$
 appears in the denominator, the SRT is enhanced (suppressed) in the anomalous
 EY-like regime. For the spin relaxation in the normal strong scattering regime, with weak HF effective magnetic
 field [$\langle |\Omega_{\rm HF}({\bf k})|\rangle \ll |\Omega|$], the condition $|\Omega+\Omega_{\rm
   HF}({\bf k})|\tau_{k,1}\ll 1$ is satisfied and hence the SRT is unchanged by
 the HF effective
 magnetic field.   

\subsubsection{All relevant scatterings}
In this part, we discuss the genuine situation with all the relevant
scatterings, which are evitable when $T=30$ K, included. The results are
plotted as red chain curves in Fig.~\ref{figyw1} with different initial spin polarizations [In Fig.~\ref{figyw1}(a), the case with $P=0.5\%$ is not shown explicitly as it is very close to the case with $P=0.1\%$]. For the transverse
spin relaxation [Fig.~1(a)], it is shown that when the electron-electron and electron-phonon scatterings are
included, with $P=0.1\%$ ($1\%$),
the SRT in Regimes I and II increases (decreases) slowly with increasing impurity density; in Regimes III and
IV the SRT remains unchanged. This is because when $T=30$
K, when the impurity density is low (high), the electron-electron (electron-impurity) scattering dominates the momentum relaxation, and hence the SRT with
low (high) impurity density is determined by the
electron-electron (electron-impurity) scattering. Specifically, in the low impurity density limit, the momentum relaxation time changes little by varying the impurity density, and hence the SRT with $P=0.1\%$ ($1\%$) varies slowly with the increase of the impurity density when $N_i\lesssim 0.01N_e$. Therefore, to observe
the different regimes of spin relaxation in the impurity density dependence clearly, one has to carry out the measurement at low
temperature with the electron-electron scattering being suppressed. The
situation with lower temperature $T=15$ K is shown Fig.~\ref{figyw1}(a) by the orange chain curve with all the relevant scatterings when
$P=0.1\%$, in which Regime II is clearly revealed.

For the longitudinal spin relaxation [Fig.~1(b)], no matter the spin polarization is small with
$P=0.1\%$ or large with $P=10\%$, when the electron-electron and electron-phonon
scatterings are included, in the anomalous EY-like regime, the SRT
decreases first slowly and then fastly with the increase of the impurity
density; in the normal strong scattering regime, the SRT is unchanged. This is
understood that the electron-electron (electron-impurity) scattering is dominant in
the momentum relaxation
when the impurity density is low (high). In Fig.~\ref{figyw1} (b), the SRT at 15 K is
also plotted by the orange chain curve with all the relevant scatterings when the polarization is parallel to the magnetic field with $P=10\%$. 

\subsection{Strong HF effective magnetic field}
\label{strong_part} 
\begin{figure}[ht]
  {\includegraphics[width=7.4cm]{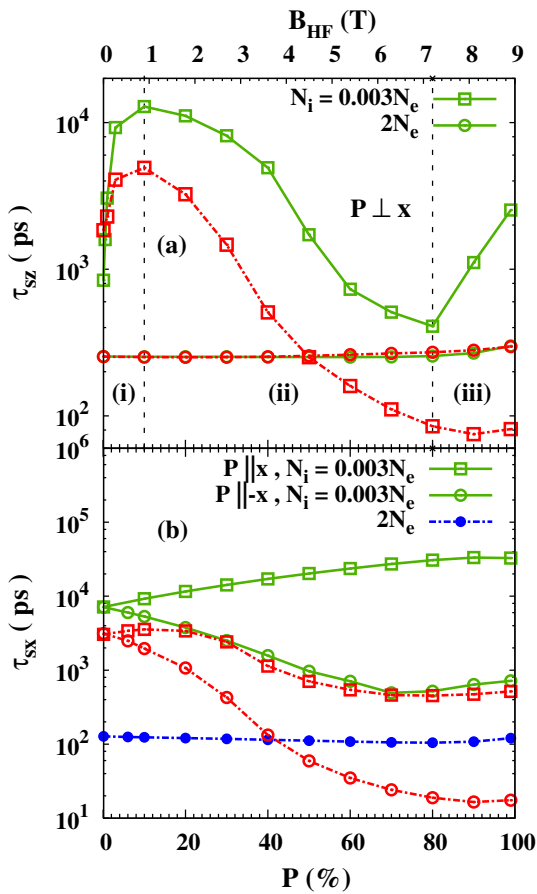}}
  \caption{(Color online) Transverse (a) and longitudinal (b)
SRTs with only the electron-impurity scattering
(green solid curves) and all the relevant scatterings (chain curves) against the
initial spin polarization with the different impurity density $N_i=0.003N_e$ and
$2N_e$. With $N_i=2N_e$, the case with ${\bf P}\parallel {\bf \hat{x}}$ is not shown explicitly as
    it is very close to the one with ${\bf P}\parallel {\bf
      -\hat{x}}$; the cases with all the
  relevant scatterings included are not shown either as they coincide with the ones
  with only the electron-impurity scattering. According to the relative magnitudes of the HF field
and the Zeeman field, for the case with $N_i=0.003N_e$ and only the electron-impurity scattering included, three regimes are shown in (a), separated by the vertical
black dashed lines: (i):
$\langle|\Omega_{\rm HF}({\bf k})|\rangle\ll |\Omega|$; (ii): $\langle|\Omega_{\rm HF}({\bf k})|\rangle\sim
|\Omega|$ and (iii): $\langle|\Omega_{\rm HF}({\bf k})|\rangle\gg
|\Omega|$. We also plot the corresponding HF field $B_{\rm HF}$ against
the initial spin poalrization (note the scale
is on the top frame of the figure).}
  \label{figyw2}
\end{figure}

In this subsection, we discuss the spin relaxation with the  strong HF effective
magnetic field [$\langle|\Omega_{\rm HF}({\bf k})|\rangle\gtrsim
|\Omega|$]. In Fig.~\ref{figyw2}, the
transverse and longitudinal SRTs are plotted against the initial spin
polarization with different impurity densities $N_i=0.003N_e$ and
$2N_e$. One notes that in the absence of the HF effective
magnetic field, when $N_i=0.003N_e$ ($2N_e$), the
transverse spin relaxation is in the anomalous DP-like (normal strong
scattering) regime [Fig.~\ref{figyw1}(a)]; the longitudinal spin relaxation is in the anomalous EY-like
(normal strong scattering) regime [Fig.~\ref{figyw1}(b)]. Hence with different
impurity density, we can discuss spin relaxation in different regimes with the
strong HF effective magnetic field by varying the initial spin polarization. Below we first
analyze the transverse and longitudinal SRTs with only the electron-impurity
scattering, in order to reveal the underlying physics for the anomalous DP relaxation with
strong HF effective magnetic field. Then we discuss the
genuine situation with all the relevant scatterings.
\subsubsection{Impurity scattering}  
In this part, we analyze the spin relaxation with only the electron-impurity
scattering. These results are plotted as green solid curves in
Fig.~\ref{figyw2}. We first analyze the transverse spin relaxation.  In
Fig.~\ref{figyw2}(a), when $N_i=0.003N_e$, it is very interesting to see that the SRT shows a peak and a basin with the increase of
the initial spin polarization. Specifically, the suppression of the SRT
  with the increase of the spin polarization in the intermediate regime is very
 intriguing and far beyond
  expectation (In conventional situation, the SRT is always enhanced due to the
  suppression of the inhomogeneous broadening by the HF field\cite{jianhua23, jianhua36,jianhua37,jianhua46}). According to the relative magnitudes of the HF effective magnetic field
[$B_{\rm HF}\equiv \langle |\Omega_{\rm HF}({\bf k})/(g\mu_B)|\rangle$] and the Zeeman magnetic
field, we further divide the spin relaxation into three regimes : (i):
$\langle|\Omega_{\rm HF}({\bf k})|\rangle\ll |\Omega|$; (ii): $\langle|\Omega_{\rm HF}({\bf k})|\rangle\sim
|\Omega|$ and (iii): $\langle|\Omega_{\rm HF}({\bf k})|\rangle\gg
|\Omega|$. When $N_i=2N_e$, the SRT increases slowly with the
increase of the initial spin polarization. Below, we analyze the situation with
$N_i=0.003N_e$ and $2N_e$, respectively.

For the transverse configuration, when $N_i=0.003N_e$,
 we first analyze the  spin relaxation in the
limit situations with
extremely weak [(i)] and strong [(iii)] HF effective magnetic fields, and then focus on
the situation with intermediate [(ii)] HF field. In Regime
(i), the HF effective magnetic field is extremely weak, 
and hence acts as a
{\em rotating} magnetic field around the Zeeman
field.\cite{Yu} The
underlying physics has been addressed in
Sec.~{\ref{weak_part}}, with the SRT determined by
Eq.~(\ref{zz_relaxation}). Accordingly, when the spin
relaxation is in the anomalous DP-like regime (when the HF field is absent) with $N_i=0.003N_e$, the SRT can be significantly enhanced by the HF effective magnetic
field. 

In Regime (iii), with $\langle|\Omega_{\rm HF}({\bf
k})|\rangle\gg |\Omega|$, the HF effective magnetic field acts as a {\em static}
magnetic field. In this situation, the system returns to the longitudinal
configuation as a large static magnetic field, i.e., the large HF effective magnetic field parallel to the spin
polarization. Accordingly, from Eq.~(\ref{xx_relaxation}), the SRT is effectively
enhanced due to the suppression of the inhomogeneous broadening by this large HF
effective magnetic field. Therefore, the SRT increases with the increase of the
initial spin polarization.

In Regime (ii), one finds that the HF effective magnetic field is in the intermediate
regime between Regimes (i) and (iii). The HF effective
magnetic field not only acts as a {\em rotating} magnetic
field around the Zeeman field, but also plays the role of a {\em static} magnetic
field. Specifically, when this effective static
magnetic field is much smaller than the Zeeman magnetic field, 
its influence to the
transverse spin relaxation can be revealed analytically. In this situation, by
noting that there exists a strong magnetic field in the plane of QWs, the
in-plane component of this effective static
magnetic field can be
neglected. Therefore, we keep the $\hat{z}$-component $B_z^{\rm HF}$ of the
effective static magnetic field in the
KSBEs, and obtain the transverse SRT when $\omega_z\equiv |g|\mu_B
B_z^{\rm HF}\ll |\Omega|$ (refer to Appendix~\ref{AA}),
{\begin{equation}
\tau_{sz}^{*-1}=\tau_{sz}^{-1}+\left\langle{\omega_{z}^2\overline{\Omega_z({\bf
        k})}^2\tau_{k,1}}/{\Omega^2}\right\rangle.
\label{extra}
\end{equation}
Hence the effective static magnetic field provides an additional spin
relaxation channel to the transverse spin relaxation. The underlying
  physics can be understood as follows. Considering the total magnetic
  field $\Omega_{\rm tot}$ from the Zeeman ($\Omega$) and effective static
  fields ($\omega_z$), when $\omega_z\ll |\Omega|$,
\begin{equation}
  \Omega_{\rm tot}=\sqrt{\Omega^2+\omega_z^2}\approx \Omega+\omega_z^2/(2\Omega).
\label{total_static}
\end{equation}
 Accordingly, the perpendicular and parallel components of the
  SOC field to $\Omega_{\rm
  tot}$ read
\begin{equation}
{\Omega_{\perp}({\bf k})\approx{\Omega_z}({\bf
    k})[1-\omega_z^2/(2\Omega^2)]}
\label{perpendicular}
\end{equation}
and
\begin{equation}
{\Omega_{\parallel}({\bf k})\approx{\Omega_z}({\bf
    k})\omega_z/\Omega},
\label{parallel}
\end{equation}
respectively. Obviously, the perpendicular component of the SOC field
[Eq.~(\ref{perpendicular})] again leads to the anomalous
DP spin relaxation and the parallel one [Eq.~(\ref{parallel})] provides the additional spin relaxation
channel [One can easily recover the additional term in
  Eq.~(\ref{extra}) by using the relation for conventional DP spin relaxation:
 $\tau_s^{-1}=\langle\overline{\Omega_{\parallel}({\bf k})}^2
  \tau_{k,1}\rangle$ (Refs.~\onlinecite{review1,Bloch,Awschalom,Zutic,fabian565,Dyakonov,wu-review,Korn})]. Accordingly, with the increase of the initial spin polarization, and
hence the effective static magnetic field, the SRT decreases.

As for the situation with $N_i=2N_e$, the transverse spin relaxation is in
the normal strong scattering regime. In this regime, the SRT can
be enhanced due to the suppression of the inhomogeneous broadening by the HF
effective magnetic field.\cite{jianhua23,jianhua36,jianhua37} Therefore, the SRT
increases with the increase of the initial spin polarization. 

We then turn to the longitudinal spin relaxation. No matter the
impurity density is low with $N_i=0.003N_e$ or high with $N_i=2N_e$, it is shown in
Fig.~\ref{figyw2}(b) that when the spin polarization is parallel
(antiparallel) to the magnetic field, the SRT increases (first decreases and then increases)
with the increase of the initial spin polarization. By noting that the spin
polarization is along the Zeeman magnetic field, the HF field can be
treated as a {\em static} effective magnetic field along the
Zeeman field. When the spin polarization is
parallel to the magnetic field, the SRT is enhanced due to the enhancement of the
total magnetic field $|\Omega+\Omega_{\rm HF}({\bf k})|$ and hence the suppression of the
inhomogeneous broadening by the HF effective magnetic field [Eq.~(\ref{xx_relaxation})]. When the spin
polarization is antiparallel to the magnetic field, if the initial spin polarization is small (large) with
$\langle|\Omega+\Omega_{\rm HF}({\bf k})|\rangle <|\Omega|$
[$\langle|\Omega+\Omega_{\rm HF}({\bf k})|\rangle >|\Omega|$], the
SRT decreases (increases) with the increase of the initial spin polarization due to the enhancement
(suppression) of the inhomogeneous
broadening by the HF effective magnetic field. 

\subsubsection{All relevant scatterings}
In this part, we analyze the genuine situation with all the relevant scatterings. 
These results are plotted as red chain curves in Fig.~\ref{figyw2}. As the electron-impurity scattering is
dominant for the situations with $N_i=2N_e$, and hence the SRT is uninfluenced by the
other scatterings, in the following we only focus on the situations with $N_i=0.003N_e$. 

For the transverse spin relaxation [Fig.~\ref{figyw2}(a)], similar to the
situation with only the electron-impurity scattering, when the spin
polarization is extremely small (large), the SRT is enhanced due to the suppression of the inhomogeneous broadening by the
extremely weak (strong) HF field; when the spin polarization is in the
intermediate regime, the SRT is suppressed due to the additional
 spin relaxation channel [Eq.~(\ref{extra})].
Moreover, the SRT can be suppressed due to the suppression of
the momentum relaxation by the electron-electron scattering.
 
Very different from the transverse configuration, some new features arise
for the longitudinal spin relaxation when the other relevant
scatterings are included. It is shown in Fig.~\ref{figyw2}(b) that when the initial spin
polarization is antiparallel to the magnetic field, similar to the
situation with only the electron-impurity scattering, the SRT also
shows a basin with the increase of the initial spin
polarization. However, in the parallel configuration, the polarization
dependence of the SRT is very different from the case with only the
electron-impurity scattering. The SRT shows a peak and a basin with the
increase of the initial spin polarization. For the parallel configuration, from
Eq.~(\ref{xx_relaxation}), one finds that on one hand, the SRT can be enhanced
due to the suppression of the inhomogeneous
broadening by the HF field; on the other hand, the SRT can be suppressed due to the
enhancement of the momentum
relaxation. In our situation, with the increase of the initial spin
polarization, the population of the electrons
  in the ${\bf k}$-space is broadened and hence the electron-electron scattering
  is enhanced. Our calculation shows that in the intermediate spin polarization
  regime, the influence of the electron-electron
  scattering to the spin relaxation is more important than the HF field, leading
  to the decrease of the SRT with the increase of the initial spin
  polarization. Moreover, for the antiparallel configuration, 
the enhancement of the
momentum relaxation also influences the spin relaxation, leading to the SRT decreasing faster than the situation
with only the electron-impurity scattering.

\section{Summary}
\label{summary}
In summary, we have investigated the role of the HF self-energy, acting as an
effective magnetic field, to the anomalous
DP spin relaxation in InAs (110) QWs under the strong magnetic
field in the Voigt configuration, whose magnitude is much larger than the
spin-orbit-coupled field. The transverse and longitudinal spin
relaxations  are considered both analytically and 
numerically. Some intriguing features beyond
expectation are revealed, which arise from the {\em unique} form of the effective inhomogeneous
broadening due to the mutually perpendicular spin-orbit-coupled field and
  strong Zeeman field.

For the
transverse configuration, we find that the behavior of the spin relaxation is
very sensitive to the HF effective magnetic field. Very different from the
conventional DP spin relaxation,\cite{jianhua36,jianhua37,jianhua46} in our
situation, even with
extremely small spin polarizations $P>
0.02\%$, the spin relaxation is significantly
influenced by the HF field. Specifically, when $P\lesssim
0.02\%$, the spin relaxation can be divided into four regimes: the normal weak scattering regime, the anomalous DP-like
regime, the anomalous EY-like regime and the normal strong scattering
regime.\cite{Zhou} However, even a slightly increase of the spin
polarization with $P=0.1\%$, these behaviors are significantly
influenced by the HF field: the
normal weak scattering regime vanishes and the {\em whole}
scattering regime lies in the strong scattering limit, with the EY-like
  behavior in the original normal weak scattering regime now contributed by the HF field. This arises from the suppression of the effective inhomogeneous broadening by the HF field. Interestingly, if one
further increases the spin polarization ($P\gtrsim 1\%$), the effective
inhomogeneous broadening is further suppressed and hence the SRT is
  divided into two rather than four regimes: the anomalous EY-like regime and the normal strong scattering
regime. These marked
influence of such small spin polarizations on the spin relaxation arises from
the fact that this unique effective inhomogeneous broadening is very small and
hence sensitive to the HF field. 

Moreover, we further find that in the spin polarization dependence, there exsits three regimes for the transverse spin
relaxation, showing a peak and a valley with
the increase of the spin polarization. This is very different from the
conventional DP spin relaxation, where the SRT increases monotonically with the
increase of the spin polarization.\cite{jianhua23,
  jianhua36,jianhua37,jianhua46}  Specifically, we reveal that the decrease of
the SRT with the increase of the spin polarization comes from the additional spin
relaxation channel provided by the HF field in
the intermediate spin polarization regime.

For the
longitudinal configuration, we find that the HF field,  acting as an effetive
static magnetic field, influences the magnitude of the total magnetic field
along the direction of the Zeeman field. No matter the initial spin
polarization is small or large, the spin relaxation is divided into two regimes: the anomalous EY-like regime and the normal strong scattering
regime. Furthermore, it is
found that the
longitudinal spin relaxation can be either suppressed or enhanced by the
HF effective magnetic field if the spin polarization is parallel or
antiparallel to the magnetic field.

\begin{acknowledgments}
This work was supported
 by the National Natural Science Foundation of China under Grant
No. 11334014, the National Basic Research Program of China under Grant No.
2012CB922002 and the Strategic Priority Research Program 
of the Chinese Academy of Sciences under Grant
No. XDB01000000.
\end{acknowledgments}

\begin{appendix}

\section{Derivation of Eq.~(\ref{extra})}
\label{AA}
We analytically derive Eq.~(\ref{extra}) based on the KSBEs
with the weak
out-of-plane static magnetic field (along the $\hat{z}$-axis) included.  When the Zeeman magnetic field satisfies
 $|\Omega|\gg \langle|\Omega_z({\bf k})+\omega_z|\rangle$, it is convenient to solve the KSBEs in the helix space.\cite{Yu} In the helix space, we further
transform the KSBEs into the interaction picture, and use the rotation wave
 and Markovian approximations.\cite{jianhua52, Haug2}

The KSBEs in the collinear space can be written as\cite{Bloch}
\begin{eqnarray}
\nonumber
&&\partial_t \rho_{\bf k}+{i}[{\Omega_z({\bf k})\sigma_z}/{2},\rho_{\bf k}]+i[{\Omega}\sigma_x/{2},\rho_{\bf
  k}]+i[{\omega_z}\sigma_z/{2},\rho_{\bf
  k}]\\
&&\mbox{}+\sum\limits_{\bf k'}W_{\bf kk'}(\rho_{\bf k}-\rho_{\bf k'})=0,
\label{KSBEs}
\end{eqnarray}
in which $W_{\bf kk'}={2\pi}|V_{\bf k-k'}|^2\delta(\varepsilon_{\bf
  k}-\varepsilon_{\bf k'})$ describes the electron-impurity scattering. For
simplicity, we retain only the linear-$k$ term in the Dresselhaus SOC, i.e., $\Omega_z({\bf k})
=\alpha k_x$ with $\alpha \equiv -\gamma_D\langle k_z^2\rangle/2$.  

 We then transform the KSBEs [Eq.~(\ref{KSBEs})] from the collinear space to the helix one by the
transformation matrix
\begin{equation}
U_{\bf k}=\left(\begin{array}{cc}
\frac{\displaystyle -\Omega}{\displaystyle \sqrt{\Omega^2+\Omega_{-}^2}} &
\frac{\displaystyle -\Omega}{\sqrt{\displaystyle \Omega^2+\Omega_{+}^2}} \\
\frac{\displaystyle \Omega_{-}}{\displaystyle
  \sqrt{\Omega^2+\Omega_{-}^2}} & \frac{\displaystyle \Omega_{+}}{\displaystyle \sqrt{\Omega^2+\Omega_{+}^2}}
\end{array}\right),
\label{UK}
\end{equation}
with $\Omega_{\pm}=(\alpha k_x+\omega_z)\pm\omega_{\rm tot}$. Here, $\omega_{\rm
  tot}=\sqrt{\Omega^2+(\alpha k_x+\omega_z)^2}$ is the total
 magnetic field.
When the strong Zeeman magnetic field satisfies $|\Omega|\gg |\alpha k_x +\omega_z|$,
Eq.\ (\ref{UK}) can be simplified into
\begin{equation}
U_{\bf k}\approx\frac{\displaystyle 1}{\displaystyle \sqrt{2}}\left(\begin{array}{cc}
A_{\bf k}-\frac{\displaystyle \alpha k_x+\omega_z}{\displaystyle 2\Omega} & A_{\bf
  k}+\frac{\displaystyle \alpha k_x+\omega_z}{\displaystyle 2\Omega}\\
A_{\bf k}+\frac{\displaystyle \alpha k_x+\omega_z}{\displaystyle 2\Omega} & -A_{\bf
  k}+\frac{\displaystyle \alpha k_x+\omega_z}{\displaystyle 2\Omega}
\end{array}\right),
\end{equation}
with $A_{\bf k}\equiv-1+(\alpha k_x+\omega_z)^2/(8\Omega^2)$. 
After the transformation, the KSBEs in the helix space become
\begin{eqnarray}
\nonumber
&&\partial_t \rho_{\bf k}^h+\frac{i}{2}\omega_{\rm tot}[\sigma_{z'},\rho_{\bf
  k}^h]+\sum\limits_{\bf k'}W_{\bf kk'}(\rho_{\bf k}^h-\rho_{\bf
  k'}^h)\\
\nonumber
&&\mbox{}+\sum\limits_{\bf k'}W_{\bf kk'}\frac{\alpha^2(k_x'-k_x)^2}{4\Omega^2}(\rho_{\bf k'}^h-\sigma_{y'}\rho_{\bf
  k'}^h\sigma_{y'})\\
&&\mbox{}+\sum\limits_{\bf k'}W_{\bf kk'}\frac{i\alpha}{2\Omega}(k_x-k_x')[\sigma_{y'},\rho_{\bf k'}^h]=0,
\label{helix}
\end{eqnarray}
with the density matrix in the helix space being
$\rho_{\bf k}^h=U_{\bf k}^{\dagger}\rho_{\bf
  k}U_{\bf k}$.

By taking the approximation 
\begin{equation}
 \omega_{\rm tot}\approx \Omega_{\rm tot}+{\alpha^2
 k_x^2}/{(2\Omega)}+{\alpha k_x\omega_z}/{\Omega}
\end{equation}
with $\Omega_{\rm tot}$ defined in Eq.~(\ref{total_static}), we then transform
the density matrix into the interaction picture as
\begin{equation}
\tilde{\rho}_{\bf k}=\exp(i\Omega_{\rm tot}\sigma_{z'} t/2)\rho_{\bf
  k}^h\exp(-i\Omega_{\rm tot}\sigma_{z'} t/2).
\end{equation}
By further defining the spin vector 
\begin{equation}
\tilde{{\bf S}}_{k}^l={\rm Tr}[\tilde{\rho}_{k,l}\bgreek{\sigma}],~~~~~\tilde{\rho}_{k,l}=\frac{1}{2\pi}\int_0^{2\pi}d\phi_{\bf k}\tilde{\rho}_{\bf k}e^{i l\phi_{\bf k}},
\end{equation}
 and applying the rotation wave approximation 
($|\Omega|\gg |\alpha k_x|$), one obtains
\begin{eqnarray}
\nonumber
&&\frac{\partial \tilde{{\bf S}}_k^{l}}{\partial t}+\frac{\Omega_{\rm so}({\bf
k})}{2\Omega\tau_{k,1}}U_1(t)\big[\tilde{{\bf S}}_k^{0}(\delta_{l1}+\delta_{l-1})-(\tilde{{\bf S}}_k^{1}+\tilde{{\bf
  S}}_k^{-1})\delta_{l0}\big]\\
&&\mbox{}+U_2(t)\tilde{{\bf S}}_k^{l}+U_3(t)(\tilde{{\bf S}}_k^{l-1}+\tilde{{\bf S}}_k^{l+1})=0
\label{matrixequation},
\end{eqnarray}
with $\Omega_{\rm so}({\bf k})\equiv\alpha k$ and $\delta_{ij}$ being the Kronecker symbol. Here, the matrices
$U_1(t)$ to $U_3(t)$ are defined by
\begin{equation}
U_1(t)=\left(\begin{array}{ccc}
0 & 0 & -\cos(\Omega_{\rm tot}t) \\
0 & 0 & -\sin(\Omega_{\rm tot}t)\\
\cos(\Omega_{\rm tot}t) & \sin(\Omega_{\rm tot}t) & 0
\end{array}\right),
\end{equation}
\begin{widetext}
\begin{equation}
U_2(t)=\left(\begin{array}{ccc}
\frac{\displaystyle 1}{\displaystyle \tau_{k,l}}+\frac{\displaystyle \Omega_{\rm
    so}^2({\bf k})}{\displaystyle 4\Omega^2\tau_{k,1}}
\delta_{l0} & \frac{\displaystyle \Omega_{\rm so}^2({\bf k})}{\displaystyle 4\Omega} & 0\\
-\frac{\displaystyle \Omega_{\rm so}^2({\bf k})}{\displaystyle 4\Omega} & \frac{\displaystyle 1}{\displaystyle \tau_{k,l}}+\frac{\displaystyle
  \Omega_{\rm so}^2({\bf k})}{\displaystyle 4\Omega^2\tau_{k,1}}
\delta_{l0} & 0\\
0 & 0 &  \frac{\displaystyle 1}{\displaystyle \tau_{k,l}}+\frac{\displaystyle
  \Omega_{\rm so}^2({\bf k})}{\displaystyle 2\Omega^2\tau_{k,1}}
\delta_{l0}
\end{array}\right),
\end{equation}
\end{widetext}
and
\begin{equation}
U_3(t)=\left(\begin{array}{ccc}
0 & \frac{\displaystyle \Omega_{\rm so}({\bf k})\omega_z}{\displaystyle 2\Omega} & 0 \\
-\frac{\displaystyle \Omega_{\rm so}({\bf k})\omega_z}{\displaystyle 2\Omega} & 0 & 0\\
0 & 0 & 0
\end{array}\right).
\end{equation}
In Eq.~(\ref{matrixequation}), by keeping terms $|l|\le 2$ and applying the
Markovian approximation, the transverse SRT is therefore
\begin{equation}
{\tau_{sz}}({\bf k})^{-1}\approx\frac{\Omega^4_{\rm so}({\bf
    k})\tau_{k,2}}{32\Omega^2}+\frac{\Omega^2_{\rm
    so}({\bf k})\tau_{k,1}}{4(1+\Omega^2\tau_{k,1}^2)}+\frac{\omega_z^2\Omega_{\rm so}^2({\bf k})\tau_{k,1}}{2\Omega^2}.
\end{equation}

\end{appendix}

\end{document}